# Interaction and Resistance: The Recognition of Intentions in New Human-Computer Interaction[*]

Vincent C. Müller

Anatolia College/ACT, Dept. of Humanities and Social Sciences,
P.O. Box 21021, 55510 Pylaia, Greece
`vmueller@act.edu`
http://www.sophia.de

**Abstract.** Just as AI has moved away from classical AI, human-computer interaction (HCI) must move away from what I call 'good old fashioned HCI' to 'new HCI' – it must become a part of cognitive systems research where HCI is one case of the interaction of intelligent agents (we now know that interaction is essential for intelligent agents anyway). For such interaction, we cannot just 'analyze the data', but we must assume intentions in the other, and I suggest these are largely recognized through resistance to carrying out one's own intentions. This does not require fully cognitive agents but can start at a very basic level. New HCI integrates into cognitive systems research and designs intentional systems that provide resistance to the human agent.

## 1  GOFAI and GOHCI

### 1.1  GOHCI "Good Old HCI"

It seems to me that there is a development in current Human-Computer Interaction research (HCI) that is analogous to developments in AI in the last 25 years or so, and perhaps there is a lesson to be learned there. Good old HCI (GOHCI) proceeded on the basis of the following image: Fairly intelligent humans interact with very stupid computers and the humans are trying to get the computers to do what the humans want – but the computers often don't get the point, so they need better 'interfaces' for the humans to tell them what to do. This is the agenda in classical GOHCI problems like 'text-to-speech' and 'speech-to-text'. These problems are structured in such a way that they cannot be solved completely, but solutions can only be approached – after all, the computers remain too stupid to get the point.

---

[*] *Nota Bene:* Before we begin, a word of caution is in order: While I am very grateful for the invitation to address an audience on human-computer interaction, this paper is a case of the blind talking about color. In the best case it can provide some theoretical ideas that might be an inspiration for future work in HCI; namely theoretical ideas from a perspective of theoretical work on artificial cognitive systems. I am grateful to the audience at the Caserta summer school on "Autonomous, Adaptive, and Context-Aware Multimodal Interfaces" (March 2010) for its encouraging response to my programmatic remarks.

### 1.2 GOFAI "Good Old Fashioned AI"

In AI, from the founding fathers onwards the basic approach, for which John Haugeland coined the expression "good old fashioned AI" or GOFAI [1] was that syntactic processing over symbolic representation is sufficient for intelligence – or perhaps even necessary, as some of the first AI theorists had claimed:

> *The Physical Symbol System Hypothesis.* A physical symbol system has the necessary and sufficient means for general intelligent action. By 'necessary' we mean that any system that exhibits general intelligence will prove upon analysis to be a physical symbol system. By 'sufficient' we mean that any physical symbol system of sufficient size can be organized further to exhibit general intelligence." [2, cf. 3]

The physical symbol system hypothesis was very good news for AI since we happen to know a system that can reproduce any syntactic processing: the computer; and we thus know that reproduction in computing machines will result in intelligence, once that reproduction is achieved. What remains are just technical problems. On this understanding, the understanding that cognitive ability in natural systems involves understanding the computational processes carried out in these systems, Cognitive Science and AI are really just two sides of the same coin. As a prominent proponent once put it: "Artificial intelligence is not the study of computers, but of intelligence in thought and action." [4]

### 1.3 Computing

It might be useful to add a little detail to the notion of 'computing' and 'syntactic' processing just introduced because they provide what I want to suggest is (or should be) the crucial difference between GOFAI and 'new AI' as well as GOHCI and 'new HCI'. The 'new' approach in both cases is the one that is not purely 'computational', not 'just the data'.

Computing in the sense used here is characterized by two features:

- It is digital (discrete-state, all relevant states are tokens of a type – a 'symbol' in the very basic sense of Newell & Simon)
- It is algorithmic (a precisely described and "effective" procedure, i.e. it definitely leads to a result). The system is thus characterized by its syntactical properties.

The Church-Turing thesis [5] adds to this that all and only the effectively computable functions can be computed by some Turing machine. This means, a programmable computer with the necessary resources (time & memory) can compute any algorithm (it is a universal Turing machine); so it is irrelevant how that computer is constructed, physically. Precisely *the same* computation can be carried out on different devices; computing is 'device independent'. Of course, the practical constraints on available resources, especially on the time needed to move through the steps of an algorithm, can mean that a problem that is theoretically computable, remains practically not 'tractable'.[1]

### 1.4 Three Levels of Description

One might object at this point that a computer is surely not just a syntactical system – after all, a computer has a size and weight, and its symbols can mean something; these are hardly syntactical properties. However, rather than asking what a computer really *is*, I think it is more fruitful to realize that a given computer can be described on several *levels*:

1. The *physical level* of the actual "realization" of the computer (e.g. electronic circuits on semiconductor devices)
2. the *syntactic level* of the algorithm computed, and
3. the *symbolic level* of content (representation), of what is computed

So, yes a computer has a size and weight (on the physical level 1) and, yes, its symbols have meaning (on the symbolic level 3), but the *computation* is on the syntactic level 2, and only there. Level 2 can be realized in various forms of level 1, i.e. the physical level does not determine which algorithm is performed (on several levels of 'algorithm'). Level 3 can be absent and is irrelevant to the function. It can have several sub-levels, e.g. the computing can symbolize sound, which is speech, which is an English sentence, which is Richard Nixon giving orders to pay blackmail money, … (levels increasing in 'thickness' in the sense to be explained presently).

Having said that, it is important to realize that the symbolic level is something that is not part of the system but attributed to it from the outside, e.g. from humans. Expressions like "The computer classifies this word as an adverb", "The computer understands this as the emotion of anger" or "The computer obeys the program" are strictly speaking nonsense – or at best metaphorical, saying that the computer carries out some program and some user *interprets* that outcome as meaningful (more details on what is only sketched here in [8]).

## 2 Intelligence and Goals

### 2.1 Thick Description

The proposed move away from GOHCI would involve moving away from this purely syntactic level, from the 'data' that we normally begin with. To illustrate what that might mean and why it might be necessary for new HCI, allow me a short deviation to social anthropology. Social anthropology analyzes human culture, so one of its concerns is what makes some observational data 'cultural' and thus relevant for the discipline. A famous proposal that Clifford Geertz developed on the basis of ideas by the philosopher Gilbert Ryle is that for humans things must somehow be *meaningful*. As an example, he considers the observational data of a quick closing movement of someone's eye. This, he says, may be part of cultural activity just if it is described 'thickly' as "a wink", rather than described 'thinly' as a "the twitch of an eye". One can wink "(1) deliberately, (2) to someone in particular, (3) to impart a particular message, (4) according to a socially established code, (5) without cognizance of the rest of the company" or (6) to parody of someone else's wink, … and so on [9]. The 'thick' description involves the meaning of the action in the culture.

---

[1] I use a classical notion of computing here, essentially derived from Turing [5] – this is not the place to defend the position that this is the only unified notion of computing we have, but some arguments in this direction can be found in [6] and [7].

This distinction between thick and thin description then allows a first characterization of culture: "As interworked system of construable signs … culture is not a power; ...; it is a context, something within which they can be intelligibly – that is thickly – described." [9]

Description of what humans do and why they do what they do must thus be "thick"; it must involve the meaning of their actions, their goals and their intentions. (It cannot be 'just the data'.) – This was missing from GOFAI and GOHCI.

## 3   HCI Is Part of Cognitive Systems Research

*Thesis I:*
HCI research is a special case of cognitive systems research

Painting the picture in a very broad brush, the move away from GOFAI is a move away from the purely syntactic system – we now recognize that physical realization plays a crucial role and that cognitive activity is embedded in bodily, emotional and volitional activity, often in interaction with other agents (the 'embodiment' of cognition).

In the light of this change, it is useful to take a second look at our explanation of HCI above, that it concerns systems that can interact with humans to achieve their goals – actually this is ambiguous: Whose goals are we talking about that of the 'system' or of the human? Since this is just a special case of the interaction that is, as we just said, crucial to cognitive activity (an aspect of embodiment), we can start to see how HCI can be a part of cognitive systems research. This is not to rehash the truism that computer systems will become 'smarter' over time, far from it: If we look at new HCI, as part of general interaction of intelligent agents, we gain a new perspective of 'thick' HCI that is not dependent on achieving 'higher levels of cognition' (whatever that means) in the machine.

## 4   Interaction Requires Resistance

Interaction of intelligent agents requires the recognition of the other agent *as* intelligent. What does that mean? Beyond a lot of little details, *intelligence* is the ability to flexibly successfully reach goals. For this reason, the recognition of intelligence requires the recognition of the agents' having goals (though not necessarily *which* goals these are). One typical way to recognize the goals of another intelligent agent is their pursuing of their goals, perhaps resisting our pursuing our own goals.

*Thesis II:*
Successful interaction requires resistance to the other agent.

Take the example of a very familiar situation: Two people walk towards each other in some narrow corridor. In order to figure out what to do, given my own goals, I might wonder: Is the other walking past me? towards me? passing on which side? One can use a rule ('walk on the right'), negotiate, but in any case one must understand the intentions – through resistance, through some force acting against my movements or intentions.

It is actually *easier* to negotiate with someone who has their own intentions than with someone who just wants to do whatever I do. In the corridor, I am more likely to collide with a super-polite person who is trying hard to get out of my way than with a bullish person who clearly indicates where he is going. It is easier to work with pedals and levers that offer resistance (sometimes called 'feedback', often an unfortunate expression); it is easier to drive a car or fly a plane if the vehicle offers a 'feel' of resistance, of 'wanting' to go this way.[2] Kicking a ball requires a certain resistance from the ball; not too much, not too little. (Note I: Even a simple example like that of the corridor always has a context. Note II: In more complex examples, the role of resistance is larger.)

Basic embodiment nicely supplements this view: our bodily actions are the basis and the model for our cognitive structure. Thus: resistance, first to the body, then to the mind. For example, we say "to grasp" when we mean to understand, and grasping requires resistance of what is grasped. In order to interact with something, we need to take it as a being with a body and a goals, accordingly something that moves purposefully.

This is one of the reasons why the notorious "Turing Test" [12], is too limited; why many people have argued that passing that test is not a sufficient condition for intelligence. But notice that it is a test of interaction, so in trying to decide whether we need new HCI or can stick to GOHCI we can take this test as a sample. In GOHCI, we are trying to pass the test without any intelligence in the system. Can this work? In new HCI, are we rejecting the test as insufficient because it offers no full verbal interaction (thus no prosody), no non-verbal interaction, no prior context, …? When we describe a situation in terms of goals and intentions, we describe it as 'meaningful', i.e. we use a 'thick' description. We need more than 'just the data'.

So, I conclude that interaction requires understanding goals and intentions, e.g. through resistance.

## 5  Resistance for New HCI

*Thesis III:* Resistance is essential even for non-cognitive HCI systems

Humans are good at interaction with other cognitive agents, like other humans or animals. We have the ability to attribute mental states to others (this is often expressed as saying we have a 'theory of mind'), even mental states that differ from my

---

[2] Caution: It is easier to know what to expect from the inexperienced because novices follow rules, experts break them [10, 11].

own – humans seem to acquire this ability around age four, as the classical 'false belief tasks' suggest. This recognition of the other as having intentional states is typically taken as a hallmark of higher-level intelligence [cf. 13]. On a neurological level, it is associated with 'mirror-neurons', the existence of which is evidence for this ability – 'seeing someone do $x$' is very closely associated with 'doing $x$'. But not only that: Humans use the very same behavior even when they know they are interacting with something that has no cognitive structure, no intentions, no goals. We are very ready to attribute 'as if' intentions to objects, to say that the car 'wants' to go left, to attribute intentions to avatars in games, to getting into a personal relation with the voice of our car navigation system – or even with the car itself (Some of this has been exploited in research that shows how humans will bond with certain 'emotional' robots; e.g. by Turkle and Dautenhahn). We assume intentions in the world, we like to use 'thick' description – one might say we are natural-born panpsychists.

This is the feature that I would suggest New HCI should exploit: Even small resistance of a computing system suggests intentions or goals (and thus intelligence) to the human user, who then knows how to interact with that system. This is not to say that New HCI should wait for computers to become sufficiently intelligent, or that they already are – far from it! What we need to do is to search for usable resistance on a basic level, on the levels that we *need* and that *we can handle* technically, at a given point in time.

Even in the long run, the aim of HCI cannot be a resistance-less computational agent. Who wants a collaborator with no aims and no desires? Someone just waiting for orders and carrying them out to the letter (needing a total description), with no desire to do well, with no initiative? The perfect intelligent collaborator has ideas, takes initiative, prevents mistakes, takes pride in their work, thinks of what I forget – she offers resistance.

Think of the small advances that have been made in certain GUI systems where one can scroll through a list by 'accelerating' that list – it is as though one accelerates an object with a weight that then continues to move on in the same direction. It needs to be accelerated against a *resistance* and it is then only slowed down by the *resistance* it encounters. (Unfortunately the cursor movement and standard object manipulation in the common graphical operating systems are not like this.) Our world is a world full of resistance, the world of resistance-less objects on standard computers is not our world.

Having said that, while scrolling down a list, pushing a lever or kicking a ball are only very loosely associated with the 'feel' of intentions through resistance, adding a little complexity to the interaction quickly does wonders. Think of how a 'Braitenberg vehicle' directly connects sensors to motor actions, and yet achieves surprisingly complex behavior [14] that seems purposeful and 'alive'. Think of how little is required for an animal or even a plant to have 'goals'.

We now know that cognitive systems research has to start with basic behavior that can be described in 'thick' terms – not just the data, and not just the human-level intelligence. New HCI would be interaction of humans with artificial agents that pursue goals and offer resistance – this is the path to successful HCI and to integration of HCI into the larger family of cognitive systems research.

777